\documentclass[11pt]{article}
\usepackage[margin=1in]{geometry}
\usepackage{amsmath,amssymb,amsthm}
\usepackage[hidelinks]{hyperref}

\newtheorem{theorem}{Theorem}
\newtheorem{lemma}{Lemma}
\newtheorem{proposition}{Proposition}
\newtheorem{corollary}{Corollary}
\newtheorem{definition}{Definition}
\theoremstyle{remark}
\newtheorem{remark}{Remark}

\newcommand{\ket}[1]{|#1\rangle}
\newcommand{\bra}[1]{\langle#1|}
\newcommand{\id}{\mathbb{I}}
\newcommand{\HC}{H_{\mathrm{C}}}
\newcommand{\Tr}{\mathrm{Tr}}

\title{Timed photon count can show deviation from classicality in coherent
state photon counting}
\author{Tahmeed Hossain\\[4pt]
\normalsize\itshape Department of Theoretical Physics, University of Dhaka,
Dhaka, Bangladesh}
\date{November 30, 2026}

\begin{document}
\maketitle

\begin{abstract}
Photon counting cannot certify any departure of coherent light from
classicality: the counting record of coherent states, however
time-resolved, is reproduced by a classical stochastic field. We first
prove a sharper, state-independent obstruction: any observables jointly
measurable with the photon numbers of two modes necessarily commute
with one another, so the complete counting-compatible data of any input
state admits a noncontextual hidden-variable model. We then attach to
each of two photodetectors a quantum clock in the sense of Stoica, a
system whose Hamiltonian generates translations and carries a
self-adjoint time operator $\hat\tau_k$. We show that $\hat\tau_k$ is
canonically conjugate, in Weyl form, to the station energy
$E_k=\HC{}_{,k}+\omega N_k$---the clock dilates the Pauli-forbidden
conjugate of the counting energy---and each detected quantum is
time-stamped by depositing its energy in the clock, so detection
conserves $E_k$. The two pairs $(\hat\tau_k,E_k)$ replace the two
canonical pairs of the continuous-variable state-independent
contextuality construction of Plastino and Cabello. The resulting
Peres--Mermin square consists of nine $\pm1$ observables that commute
with both photon counts and are conserved by the counting dynamics, yet
attain $\chi=6>4$ identically, for every input state and arbitrary,
even thermal, clock states, while the counting marginals remain exactly
Poissonian and undisturbed. The timed photon-count record of coherent
light thus deviates maximally from noncontextual classicality, while
the untimed record never can. Ideal clocks are not required: a cyclic
$M$-level register, minimally one qubit per detector, realizes the
square exactly; hard-wall registers converge exponentially; and one
station already suffices algebraically. All operator identities are
verified numerically.
\end{abstract}

\section{Introduction}\label{sec:intro}

A single-mode coherent state $\ket{\alpha}$ is the standard model of ideal
laser light, and it is also the standard example of a state whose
photodetection statistics are completely classical. Its Glauber--Sudarshan
$P$ function is a point measure \cite{Glauber1963,Sudarshan1963}, so every
normally ordered correlation function, and therefore every photocounting
distribution in the Kelley--Kleiner--Mandel theory of photodetection
\cite{KelleyKleiner1964,MandelWolf}, coincides with that of a classical
field of fixed complex amplitude: the counts are Poissonian, the Mandel
parameter vanishes, $g^{(2)}=1$, and the time-resolved click record is a
Poisson point process. Nonclassicality of light is accordingly defined by
the failure of the $P$ function to be a probability measure
\cite{TitulaerGlauber1965}, with sub-Poissonian counting as its
operational flag \cite{Mandel1979}. For coherent states, photon
counting---however refined---sees nothing nonclassical.

Contextuality offers a different notion of nonclassicality, attached to
observable algebras rather than to states. The Bell--Kochen--Specker
theorem \cite{Bell1966,KochenSpecker1967} excludes noncontextual value
assignments to quantum observables, and the Peres--Mermin (PM) square
\cite{Peres1990,Mermin1990} turns the exclusion into a state-independent
inequality: nine dichotomic observables, arranged in six measurement
contexts, whose correlations reach $\chi=6$ in every quantum state while
every noncontextual model obeys $\chi\le4$ \cite{Cabello2008}.
State-independent contextuality (SIC) has been verified with ions,
photons, and neutrons \cite{Kirchmair2009,Amselem2009}, and is reviewed in
Ref.~\cite{BudroniRMP}. Plastino and Cabello (PC) extended
state-independent contextuality to continuous variables
\cite{PlastinoCabello2010}, building the square from modular functions
\cite{APP1969} of two canonical pairs $(x_1,p_1)$,
$(x_2,p_2)$---eighteen Hermitian observables, paired into nine
unit-modulus complex ones---a structure anticipated by Clifton
\cite{Clifton2000} and developed into general phase-space and
modular-variable constructions in
Refs.~\cite{Asadian2015,LaversanneFinot2017}.

These two notions of nonclassicality do not meet in a photon-counting
experiment, and the first result of this paper is that they cannot. All
continuous-variable contextuality constructions we are aware
of---quadrature displacements
\cite{PlastinoCabello2010,Asadian2015,LaversanneFinot2017}, pseudospin or
parity-based squares \cite{Chen2002,McKeown2011}, displaced-parity
correlations \cite{BanaszekWodkiewicz1999}---use observables that fail to
commute with the photon number: measuring them is incompatible with
keeping the counting record. This is not an accident of the known
constructions. Proposition~\ref{prop:nogo} below shows that the commutant
of the pair $(N_1,N_2)$ on two-mode Fock space is abelian: \emph{any}
family of observables jointly measurable with both photon counts is
mutually commuting, possesses a joint probability distribution in every
state, and therefore admits a noncontextual model outright.
Counting-compatible data cannot violate any noncontextuality inequality,
for any state of the light. For coherent states the classicality is
twofold: the counting distribution itself is in addition reproduced by a
classical stochastic field. In this precise sense, photon counting alone
cannot distinguish a coherent state from classical statistics.

The obstruction has the same origin as Pauli's theorem \cite{Pauli1933}:
the number operator is discrete and bounded below, so no self-adjoint
operator on the mode's Hilbert space is canonically conjugate to
$\omega N$; the candidates are periodic phases
\cite{SusskindGlogower1964,PeggBarnett1989} or non-Weyl constructions on
dense domains \cite{Galapon2002}. The second half of the paper shows that
the obstruction is removed---and the abelian no-go turned into a maximal,
state-independent violation---by giving the detector a quantum clock. We
use the translational quantum system (TQS) of Stoica \cite{Stoica2024}: a
system whose Hamiltonian $\HC=-i\partial_\tau$ generates translations,
with spectrum $\mathbb{R}$ and a self-adjoint time operator $\hat\tau$
conjugate to it. Attaching one TQS to each of two detection stations and
defining the station energy $E_k=\HC{}_{,k}+\omega N_k$, we prove that
$(\hat\tau_k,E_k)$ is a genuine Weyl pair (Lemma~\ref{lem:weyl}): the
clock dilates the forbidden conjugacy, with the photon number reappearing
as the multiplicity label of the energy representation. The dictionary
\[
(x_k,\,p_k,\,p_0)\;\longmapsto\;(\hat\tau_k,\,E_k,\,\omega)
\]
then transcribes the PC construction verbatim. Section~\ref{sec:square}
builds the witness in that order: first the primary two-mode, two-station
square---in a dichotomic form (nine $\pm1$ observables;
Theorem~\ref{thm:square}) and then in the original PC form of nine
unitaries, i.e.\ eighteen Hermitian observables
(Proposition~\ref{prop:pcform})---and only afterwards shows that a single
station already suffices algebraically (Proposition~\ref{prop:digits}),
with an explicit answer to the natural question of how \emph{two}
commuting qubit pairs can be extracted from \emph{one} canonical pair.
The square satisfies the operator identity $\hat\chi=6\,\id$: the
violation is maximal for every state of modes and clocks, with no clock
preparation required. The parity of the energy cell
$\lfloor E_k/\omega\rfloor$ factorizes as (clock-bin parity) $\times$
(photon-number parity) $(-1)^{N_k}$, so the witness consumes the very
click record that looks classical on its own.

Two further results tie the construction to detection physics rather than
mere algebra (Sec.~\ref{sec:detection}). First, every observable of the
square commutes with both photon counts, so each context is jointly
measurable with the counts and leaves every counting marginal untouched;
for coherent inputs the counts stay exactly Poissonian while the same
runs violate noncontextuality maximally. Second, modeling detection by
the energy-conserving absorption coupling
$V_k=\lambda_k(a_kT_k+a_k^\dagger T_k^\dagger)$, in which each absorbed
quantum $\omega$ is deposited in the clock---the clock time-stamps the
event---all nine observables are conserved, term by term and jump by
jump, by the detection dynamics and by the associated counting
semigroup: the witness can be read before, during, or after the
accumulation of the click record, with the same statistics. The analogous
clock-only observables \emph{anticommute} with $V_k$ and are scrambled by
every detection event, while mode-only observables are excluded by the
no-go: the square sits precisely in the intersection selected by energy
conservation of time-stamped counting. The modes are not spectators; the
$Z$-type outcomes are digit functions of (clock bin $+$ photon count).

Finally, ideal clocks are not needed (Sec.~\ref{sec:physclocks}). A
cyclic $M$-level energy register with level spacing $\omega/s$ realizes
the entire construction exactly whenever $2s\mid M$---minimally, $M=2s$:
for $s=1$, one qubit register per detector---and a hard-wall register of
$K$ levels reaches $\chi=6$ up to corrections that vanish exponentially
in $K$. All algebraic identities and the finite-size table are verified
numerically to machine precision (Appendix~\ref{app:numerics}).

A caveat on interpretation is owed up front. State-independent
contextuality is a property of an observable set, not of a state, so the
violation reported here does not certify nonclassicality of coherent
light in the $P$-function sense, nor does it contradict the classical
simulations and classical-light experiments of
Refs.~\cite{Frustaglia2016,Markiewicz2019,ZhangKim2019,Qu2021}; the
relation to these works, and to the equivalence between contextuality and
Wigner negativity for Gaussian-type measurement models \cite{Booth2022},
is discussed in Sec.~\ref{sec:coherent}. What the construction
establishes is a dichotomy about the measurement side of photodetection:
the untimed photon-count record is noncontextual for every state
(Proposition~\ref{prop:nogo}), while the \emph{timed} photon-count
record---the same counts, read jointly with a quantum time
reference---deviates maximally and state-independently from noncontextual
classicality (Theorem~\ref{thm:square}), at no cost to the counting
statistics (Theorem~\ref{thm:conserve}). That is the sense of the title.

The paper is organized as follows. Section~\ref{sec:nogo} formalizes
counting-compatible measurements and proves the abelian no-go.
Section~\ref{sec:clock} recalls Stoica's TQS, sets up the two stations,
and proves the clock dilation lemma. Section~\ref{sec:square} constructs
the two-station witness (dichotomic and PC forms) and then the
single-station variant. Section~\ref{sec:detection} proves compatibility
with the detection process and gives the operational protocol.
Section~\ref{sec:coherent} spells out the coherent-state phenomenology
and the scope of the claims. Section~\ref{sec:physclocks} treats finite
and physical clocks. Section~\ref{sec:discussion} compares with prior
work and lists open problems. Proofs are collected in
Appendix~\ref{app:proofs}; Appendix~\ref{app:numerics} documents the
numerical validation. We set $\hbar=1$ and, for notational simplicity,
give both stations the same frequency $\omega$; every station-local
definition below generalizes verbatim to $\omega\to\omega_k$, and no
observable mixes the two stations' frequencies.

\section{Photon counting is noncontextual by itself}\label{sec:nogo}

Let $\mathcal{F}=\mathcal{F}_1\otimes\mathcal{F}_2$ be the Fock space of
two modes $a_1,a_2$ with number operators $N_k=a_k^\dagger a_k$. The
joint eigenbasis $\ket{n_1,n_2}$ is nondegenerate. Everything in this
section refers to $\mathcal{F}$ alone; ancillas enter from
Sec.~\ref{sec:clock} on, and escaping the following proposition is
exactly what they are for.

\begin{definition}[Counting-compatible observables]\label{def:cc}
A bounded operator $O$ on $\mathcal{F}$ (or on
$\mathcal{K}\otimes\mathcal{F}$ for an ancilla space $\mathcal{K}$) is
\emph{counting-compatible} if $[O,N_1]=[O,N_2]=0$. A POVM is
counting-compatible if all its effects are.
\end{definition}

The definition captures every measurement whose statistics can be
gathered jointly with, and without disturbing, number-resolved photon
counting at the two stations. On $\mathcal{F}$ it includes all functions
of the click record of direct (phase-insensitive) detection: in the
continuous photodetection theory of Srinivas and Davies
\cite{SrinivasDavies1981}, the probability of any click record of free
modes is a diagonal functional of the state in the number basis, i.e.\
the statistics of effects $f(N_1,N_2)$; time stamps drawn from a
\emph{laboratory} clock are classical parameters attached to the jumps
and do not leave this class. (Homodyne and heterodyne schemes leave the
class, because adding a local oscillator replaces counting of the signal
mode by counting of mixed modes; we return to this in
Sec.~\ref{sec:discussion}.)

\begin{proposition}[Counting data admits a noncontextual model]\label{prop:nogo}
On $\mathcal{F}$, every counting-compatible bounded operator is a
function of $(N_1,N_2)$. Consequently:
\begin{enumerate}
\item[(i)] any family of counting-compatible observables on $\mathcal{F}$
is mutually commuting and possesses, in every state $\rho$, a joint
probability distribution reproducing the statistics of all its
(compatible) subsets;
\item[(ii)] no noncontextuality inequality can be violated by
counting-compatible measurements on $\mathcal{F}$, for any state; and
\item[(iii)] no nine counting-compatible operators on $\mathcal{F}$ can
satisfy the Peres--Mermin product relations
[Eq.~\eqref{eq:pmrelations} below].
\end{enumerate}
\end{proposition}

\begin{proof}
If $[O,N_k]=0$ for $k=1,2$ then $O$ preserves each joint eigenspace of
$(N_1,N_2)$; these are one-dimensional, so
$O=\sum_{n_1,n_2}o_{n_1n_2}\ket{n_1,n_2}\bra{n_1,n_2}$ is a function of
$(N_1,N_2)$. (i)~All such operators are simultaneously diagonal; the
projective measurement of $(N_1,N_2)$ refines every counting-compatible
measurement, and the distribution
$p(n_1,n_2)=\bra{n_1,n_2}\rho\ket{n_1,n_2}$, pushed forward through the
outcome functions, is the required joint distribution. (ii)~The hidden
variable $\lambda=(n_1,n_2)$, distributed as $p$, assigns to every
counting-compatible observable the deterministic value $o_{n_1n_2}$,
independently of the context in which it is measured; this is a
noncontextual model for all such data. (iii)~For mutually commuting
operators the product of the three row products equals the product of
the three column products (both equal the product of all nine factors),
so the PM sign pattern---five contexts with product $+\id$ and one with
$-\id$---is contradictory.
\end{proof}

\begin{corollary}[Two-layer classicality of counting coherent light]
\label{cor:twolayer}
For any input state, counting-compatible data admits a noncontextual
model [layer 1]. If in addition the input is a coherent state (or any
state with a nonnegative $P$ function), the distribution $p(n_1,n_2)$
itself is that of a classical stochastic field: independent Poisson
counts for $\ket{\alpha_1}\otimes\ket{\alpha_2}$
\cite{Glauber1963,Sudarshan1963,KelleyKleiner1964,MandelWolf} [layer 2].
No functional of photon-counting data, however time-resolved by
laboratory clocks, can distinguish such light from a classical ensemble.
\end{corollary}

Proposition~\ref{prop:nogo} fixes the rules of the game: any
contextuality visible ``in the photon count'' must come from enlarging
the system by an ancilla whose coupling to the counting process is
physically motivated, while keeping every measured observable
counting-compatible on the enlarged space. A clock attached to the
detector is the natural such ancilla, for a reason the next section makes
exact: what counting lacks is the conjugate of the energy it counts.
\section{A quantum clock at each detector}\label{sec:clock}

\subsection{Translational quantum systems}

Pauli's argument \cite{Pauli1933} forbids a self-adjoint time operator
canonically conjugate (in Weyl form) to a Hamiltonian whose spectrum is
bounded below or discrete: conjugacy makes the unitaries $e^{ibT}$ shift
the energy spectrum by arbitrary amounts, so the spectrum must be
$\mathbb{R}$. The standard responses either abandon self-adjointness
(covariant POVMs \cite{Busch2008}), abandon the Weyl form (dense-domain
canonical pairs \cite{Galapon2002}), or accept periodicity (phase
operators conjugate to number
\cite{SusskindGlogower1964,PeggBarnett1989}). Stoica's proposal
\cite{Stoica2024,Stoica2022} takes the remaining branch: admit a clock
whose Hamiltonian genuinely generates translations. A \emph{translational
quantum system} (TQS) is a system with Hilbert space
$\mathcal{K}=L^2(\mathbb{R})$ (times a possible multiplicity space) on
which
\begin{equation}
\HC=-i\,\partial_\tau,\qquad (\hat\tau\,\psi)(\tau)=\tau\,\psi(\tau),\qquad
[\hat\tau,\HC]=i\,\id,
\label{eq:tqs}
\end{equation}
so that both $\hat\tau$ and $\HC$ have spectrum $\mathbb{R}$ and the pair
is unitarily equivalent to the Schr\"odinger pair $(\hat q,\hat p)$.
Reference \cite{Stoica2024} argues that such systems are empirically
adequate---ideal measurements, standard clocks, and quantum
representations of time-reversible deterministic dynamics already contain
them, and unitarity forbids the naive runaway decay that an
unbounded-below spectrum suggests---and shows that composing a generic
system with a TQS clock makes the \emph{total} Hamiltonian a generator of
translations, conjugate to the clock's time operator. A variant with
positive energy and a modified intrinsic generator appears in
Ref.~\cite{Stoica2026}. We take Eq.~\eqref{eq:tqs} as the definition of
the ideal clock and return to bounded, finite-level substitutes in
Sec.~\ref{sec:physclocks}; for the relational (Page--Wootters) reading of
such clocks see
Refs.~\cite{PageWootters1983,GLM2015,SmithAhmadi2019,HSL2021,Chataignier2024}.

\subsection{Setting and the dilation lemma}

Our system consists of two detection stations $k=1,2$ at two locations.
Station $k$ carries one field mode $a_k$ of frequency $\omega$ with free
Hamiltonian $\omega N_k$, and one TQS clock $(\hat\tau_k,\HC{}_{,k})$
attached to the detector. The station Hilbert space is
$\mathcal{H}_k=\mathcal{K}_k\otimes\mathcal{F}_k
=L^2(\mathbb{R})\otimes\mathcal{F}_k$, and the total space is
$\mathcal{H}=\mathcal{H}_1\otimes\mathcal{H}_2$. The central object is
the \emph{station energy}
\begin{equation}
E_k \;=\; \HC{}_{,k}+\omega N_k ,
\label{eq:stationenergy}
\end{equation}
the sum of the clock energy and the counting energy, which is essentially
self-adjoint on the natural domain and has absolutely continuous spectrum
$\mathbb{R}$.

\begin{lemma}[Clock dilation of the number operator]\label{lem:weyl}
On $\mathcal{H}_k$, the pair $(\hat\tau_k,E_k)$ satisfies the Weyl
relations
\begin{equation}
e^{ia\hat\tau_k}\,e^{ibE_k}\;=\;e^{-iab}\,e^{ibE_k}\,e^{ia\hat\tau_k},
\qquad a,b\in\mathbb{R},
\label{eq:weylrel}
\end{equation}
and there is a unitary
$W:\mathcal{H}_k\to L^2(\mathbb{R})\otimes\ell^2(\mathbb{N})$ with
\begin{equation}
W\hat\tau_k W^\dagger=\hat q\otimes\id,\qquad
W E_k W^\dagger=\hat p\otimes\id,\qquad
W N_k W^\dagger=\id\otimes\,\mathrm{diag}(n).
\label{eq:dilation}
\end{equation}
In particular $(\hat\tau_k,E_k)$ is a canonical pair with both spectra
$\mathbb{R}$, and the photon number is the multiplicity label of the
energy representation.
\end{lemma}

The proof (Appendix~\ref{app:proofs}) is an explicit change of variables
$(\varepsilon,n)\mapsto(e,n)$ with $e=\varepsilon+\omega n$ in the
clock-energy representation. Two remarks delimit what the lemma does and
does not say.

\begin{remark}[Pauli bookkeeping]\label{rem:pauli}
$\hat\tau_k$ commutes with $N_k$; it is \emph{not} conjugate to
$\omega N_k$, which is impossible, but to $E_k$, whose spectrum the clock
unrolls to $\mathbb{R}$. All count dependence of the observables below
enters through $E_k-\HC{}_{,k}=\omega N_k$; concretely, modular functions
of $E_k$ factor into clock functions times photon-number parities
[Eq.~\eqref{eq:factorize}]. The clock carries the conjugacy; the mode
contributes the discrete part that counting reads out.
\end{remark}

\begin{remark}[Laboratory time stamps are not clocks]\label{rem:labtime}
Recording the laboratory time of each click adds a classical parameter to
the record and keeps all effects inside the abelian algebra of
Proposition~\ref{prop:nogo}. The TQS replaces the parameter by an
operator $\hat\tau_k$---a dynamical, quantum time reference local to the
detector---and it is only the operator-valued reference that supports
observables outside the commutant of $(N_1,N_2)$ while remaining
counting-compatible.
\end{remark}

\section{The witness}\label{sec:square}

This section is ordered from the physical to the minimal: first the
per-station toolkit (Sec.~\ref{sec:stationobs}), then the primary
construction---two modes, two stations, two clock--mode pairs---in its
dichotomic form (Sec.~\ref{sec:twostation}) and in the original
Plastino--Cabello form of nine unitaries, i.e.\ eighteen Hermitian
observables (Sec.~\ref{sec:pcform}); only then do we show that one
station is already enough algebraically, and explain exactly how two
commuting qubit pairs arise from a single canonical pair
(Sec.~\ref{sec:onestation}).

\subsection{Station observables}\label{sec:stationobs}

Fix a station $k$ and drop its index. With $E$ as in
Eq.~\eqref{eq:stationenergy}, define the unitaries
\begin{equation}
T \;=\; e^{i\omega\hat\tau},\qquad
S \;=\; e^{i\pi E/\omega},
\label{eq:TS}
\end{equation}
and, writing $c(E)=\lfloor E/\omega\rfloor$ for the energy cell (a Borel
function of $E$; cell boundaries have measure zero on the absolutely
continuous spectrum), the spectral projectors and dichotomic observables
\begin{equation}
P^{\pm}=\chi_{\pm}\bigl(c(E)\bigr),\qquad
Z=P^{+}-P^{-}=(-1)^{c(E)},\qquad
X=T P^{+}+T^{\dagger}P^{-},\qquad
Y=iXZ,
\label{eq:XYZ}
\end{equation}
where $\chi_+$ ($\chi_-$) selects even (odd) cells. $T$ raises the clock
energy, hence $E$, by one photon quantum:
$T\ket{\varepsilon}=\ket{\varepsilon+\omega}$, $T^\dagger ET=E+\omega$.
Because
$\lfloor(\varepsilon+\omega n)/\omega\rfloor
=\lfloor\varepsilon/\omega\rfloor+n$, the $E$-modular objects factorize
into clock and count parts,
\begin{equation}
Z=(-1)^{\lfloor \HC/\omega\rfloor}\otimes(-1)^{N},
\qquad
S=e^{i\pi \HC/\omega}\otimes(-1)^{N}:
\label{eq:factorize}
\end{equation}
\emph{the photon-number parity sits inside every $Z$- and $S$-type
observable}. Operationally, a $Z$ measurement is a number-resolved count
$n$ together with a reading of the clock-energy bin
$j=\lfloor\varepsilon/\omega\rfloor$, with outcome $(-1)^{j+n}$:
counting, aided by the clock.

\begin{lemma}[Station algebra]\label{lem:station}
On $\mathcal{H}_k$: $T P^{\pm}=P^{\mp}T$ and $ST=-TS$; $X$, $Y$, $Z$ are
Hermitian unitary involutions with
\[
\{X,Z\}=\{X,Y\}=\{Y,Z\}=0,\qquad XZY=-i\,\id,
\]
and every one of $T,S,X,Y,Z,P^\pm$ commutes with $N$. Consequently
$\mathcal{H}_k\cong\mathbb{C}^2\otimes\mathcal{H}_k'$ with
$(X,Y,Z)\cong(\sigma_x,\sigma_y,\sigma_z)\otimes\id$: each station
carries a \emph{cell-parity qubit} of its conserved energy, jointly
measurable with its photon count.
\end{lemma}

Proofs are in Appendix~\ref{app:proofs}. Note what would fail without the
clock: on the mode alone, $e^{i\pi N}$ exists but no unitary conjugate
shift does (the number shift is a nonunitary isometry
\cite{SusskindGlogower1964}), and Proposition~\ref{prop:nogo} excludes
any counting-compatible anticommuting partner for $(-1)^N$.

\subsection{The primary witness: two modes, two stations}
\label{sec:twostation}

Arrange the two stations' triples in the Peres--Mermin pattern
\begin{equation}
\mathcal{S}\;=\;
\begin{array}{|c|c|c|}
\hline
X_1 & X_2 & X_1X_2\\
\hline
Z_2 & Z_1 & Z_1Z_2\\
\hline
X_1Z_2 & Z_1X_2 & Y_1Y_2\\
\hline
\end{array}
\label{eq:square}
\end{equation}
with the six contexts given by the three rows $R_1,R_2,R_3$ and three
columns $C_1,C_2,C_3$ (ordered products of the three entries). All nine
entries are Hermitian involutions; entries within each row and each
column commute pairwise; and
\begin{equation}
R_1=R_2=R_3=C_1=C_2=+\id,\qquad C_3=-\id.
\label{eq:pmrelations}
\end{equation}

\begin{theorem}[State-independent violation, clocks included]\label{thm:square}
Define $\hat\chi=R_1+R_2+R_3+C_1+C_2-C_3$. Then, as an operator identity
on $\mathcal{H}$,
\begin{equation}
\hat\chi=6\,\id,
\end{equation}
so $\langle\chi\rangle_\rho=6$ for every state $\rho$ of modes and
clocks, pure or mixed---in particular for coherent light
$\ket{\alpha_1}\otimes\ket{\alpha_2}$ with arbitrary, e.g.\ thermal,
clock states. Every noncontextual model obeys $\chi\le4$. All nine
observables are counting-compatible, $[O,N_1]=[O,N_2]=0$.
\end{theorem}

The proof (Appendix~\ref{app:proofs}) is the Mermin algebra of
Lemma~\ref{lem:station} plus the standard parity argument for the bound
\cite{Mermin1990,Cabello2008}; counting-compatibility is manifest from
Lemma~\ref{lem:station}. No clock preparation, phase stabilization, or
state assumption enters anywhere: the clocks need not even be
initialized.

At the level of outcomes, context $R_2=\{Z_2,Z_1,Z_1Z_2\}$ is measured by
counting $n_1,n_2$ and reading the two clock bins $j_1,j_2$; the outcomes
$(-1)^{j_k+n_k}$ are individually random for a generic state, but their
product is fixed by the third reading with certainty, in every state.
$X$-type observables require interferometry between adjacent energy
cells of the station (a Ramsey-type measurement of the clock line,
conditioned on the count; the joint projective measurement exists because
$[X_k,N_k]=0$). Rows $R_1,R_2$ and columns $C_1,C_2$ are products of
station-local measurements; contexts $R_3$ and $C_3$ involve joint
eigenbases entangled across the two stations, as in every bipartite
reading of the PM square \cite{Aravind2002}---this is where the two
locations must be brought together (or pre-shared entanglement spent),
and we make no locality claim; the test is of noncontextuality.

\subsection{The Plastino--Cabello form: nine unitaries, eighteen
observables}\label{sec:pcform}

The original PC construction \cite{PlastinoCabello2010} uses eighteen
Hermitian observables, paired into nine unit-modulus complex observables
$U=U'+iU''$ (with $U',U''$ commuting Hermitian operators), built from
Weyl exponentials of two canonical pairs. Under the dictionary
$(x_k,p_k,p_0)\mapsto(\hat\tau_k,E_k,\omega)$ it transcribes verbatim
onto the two clock--mode pairs:

\begin{proposition}[PC transcription]\label{prop:pcform}
The nine unitaries
\begin{equation}
\begin{array}{lll}
A=T_1, & B=S_2, & C=T_1^\dagger S_2^\dagger,\\[2pt]
a=T_2^\dagger, & b=S_1, & c=T_2S_1^\dagger,\\[2pt]
\alpha=T_2T_1^\dagger, & \beta=S_1^\dagger S_2^\dagger, &
\gamma=T_1S_1T_2^\dagger S_2,
\end{array}
\label{eq:pcops}
\end{equation}
commute pairwise within each row and each column of their PM
arrangement, and satisfy
$ABC=abc=\alpha\beta\gamma=Aa\alpha=Bb\beta=+\id$ and $Cc\gamma=-\id$.
Each is a single Weyl exponential of $(\hat\tau_k,E_k)$ up to
Baker--Campbell--Hausdorff phases; e.g.\
$\gamma=e^{i(\omega\hat\tau_1+\pi E_1/\omega)}\,
e^{-i(\omega\hat\tau_2-\pi E_2/\omega)}$. The real and imaginary parts of
the nine unitaries are PC's eighteen Hermitian observables, here concrete
functions of the clock times and station energies. Every entry commutes
with $N_1$ and $N_2$, and the $S$-type entries carry the photon-number
parities through Eq.~\eqref{eq:factorize}.
\end{proposition}

For this unitary form, PC derived the noncontextual bound
$3\sqrt3\approx5.19$ against the quantum value $6$ under a unit-modulus
assumption on the assigned values; Asadian \emph{et al.}\
\cite{Asadian2015} sharpened the analysis for general noncontextual
models. We quote both and rely on neither: the dichotomic square of
Theorem~\ref{thm:square}, with its standard bound $4$, is our primary
object, and the transcription is recorded to make contact with the
continuous-variable literature. Appendix~\ref{app:proofs} verifies the
context relations of Eq.~\eqref{eq:pcops} directly.

\subsection{One station suffices algebraically}\label{sec:onestation}

The two-station setting is the physical one---two detectors at two
locations, local counts, four of the six contexts products of
station-local measurements---and it matches the two-pair structure of
the CV literature. It is natural to ask whether it is also
\emph{necessary}. It is not, and seeing why sharpens what the resource
really is.

What the PM square needs is not two canonical pairs but \emph{two
commuting copies of the Pauli algebra}: two triples of anticommuting
involutions, each triple commuting elementwise with the other. PC
obtained one Pauli copy per canonical pair because they restricted
themselves to Weyl (displacement) operators $e^{iux+ivp}$, whose
commutation phases are rigid; within that restricted class the
restriction is essential---Asadian \emph{et al.}\ proved that a PM
square of displacement operators cannot live on a single mode
\cite{Asadian2015}. But a single canonical pair acts irreducibly on the
\emph{infinite-dimensional} space $L^2(\mathbb{R})$, whose bounded
operators contain infinitely many commuting qubit factors; the
obstruction is specific to displacement operators, not to single
systems. Conditional shifts---operators of the form
$T^{m}P+T^{\dagger m}P^{\perp}$, which lie outside the Weyl
group---extract those qubits explicitly:

\begin{proposition}[Energy-digit square on one station]\label{prop:digits}
On a single station (one clock, one mode), write the cell number in
binary,
\[
c(E)=d_0+2d_1+4r,\qquad d_0,d_1\in\{0,1\},\quad r\in\mathbb{Z},
\]
with $d_0=c\bmod2$, $d_1=\lfloor c/2\rfloor\bmod2$,
$r=\lfloor c/4\rfloor$, and let $Q^{(i)}_v$ project onto $d_i=v$. Define
\[
Z_A=(-1)^{d_0},\quad X_A=T\,Q^{(0)}_0+T^\dagger Q^{(0)}_1,\qquad
Z_B=(-1)^{d_1},\quad X_B=T^2Q^{(1)}_0+T^{\dagger2}Q^{(1)}_1 .
\]
Then, in the coordinates $(d_0,d_1,r,n)$,
$X_A=\sigma_x^{(d_0)}\otimes\id$ and $X_B=\sigma_x^{(d_1)}\otimes\id$
exactly: $(X_A,Z_A)$ and $(X_B,Z_B)$ generate two commuting copies of
the Pauli algebra on one conjugate pair $(\hat\tau,E)$. The PM square
built on them satisfies $\hat\chi=6\,\id$, with all nine observables
commuting with the single count $N$ and conserved by the detection
coupling $V_k$ of Sec.~\ref{sec:detection}.
\end{proposition}

The mechanism deserves spelling out, since a bare shift would fail. The
unit shift $c\mapsto c+1$ always flips $d_0$, but when $d_0=1$ it also
carries into $d_1$ ($1+1=10$ in binary): a plain $T$ is \emph{not} a
$\sigma_x$ on the lowest digit. Conditioning the \emph{direction} of the
shift on the digit removes the carry entirely: on $d_0=0$ shift up
($c\mapsto c+1$: digit $0\to1$, no carry), on $d_0=1$ shift down
($c\mapsto c-1$: digit $1\to0$, no borrow)---so $X_A$ toggles $d_0$ and
touches nothing else. Likewise $X_B$ shifts by two cells with direction
conditioned on $d_1$, toggling $d_1$ alone. Each conditional shift is
built only from clock translations and spectral projectors of $E$, so it
commutes with $N$ and with every jump of the detection dynamics; the
proof, and the carry bookkeeping for negative $c$, are in
Appendix~\ref{app:proofs}, and all identities are verified numerically.

Thus one timed counter is already maximally contextual; the second
station buys operational, not algebraic, content (local contexts,
separated counts, and the literal PC correspondence). The price of the
one-station version is practical: its two qubits require clock
interferometry at two binary scales, shifts by $\omega$ and by
$2\omega$, on the same conserved energy.
\section{Compatibility with the detection process}\label{sec:detection}

So far the clock is kinematics. The physics of the proposal is that the
same clock can serve as the detector's internal energy register,
time-stamping each absorption, and that the square is precisely the
structure this dynamics conserves. Model detection at station $k$ by the
energy-conserving absorption coupling
\begin{equation}
V_k=\lambda_k\bigl(a_k T_k+a_k^\dagger T_k^\dagger\bigr),
\qquad [E_k,V_k]=0 :
\label{eq:Vk}
\end{equation}
each absorbed photon deposits its quantum $\omega$ in the clock ($a_k$
lowers $\omega N_k$ by $\omega$, $T_k$ raises $\HC{}_{,k}$ by $\omega$).
Explicitly, on eigenstates,
$a_kT_k\,(\ket{\varepsilon}_{\mathrm{C}}\otimes\ket{n})
=\sqrt{n}\,\ket{\varepsilon+\omega}_{\mathrm{C}}\otimes\ket{n-1}$: the
exchange is resonant, $[H_0,V_k]=0$ with $H_0=\sum_kE_k$, and among the
kinematically available conjugate partners $\HC{}_{,k}+\lambda N_k$ of
$\hat\tau_k$ it is exactly $\lambda=\omega$, i.e.\ $E_k$, that the
dynamics conserves. This is a Jaynes--Cummings-type exchange with the
clock ladder in place of the two-level atom, and the natural jump
operator for a quantum-jump (counting) unraveling \`a la
Srinivas--Davies \cite{SrinivasDavies1981} is
$L_k=\sqrt{\gamma_k}\,a_kT_k$: a click is an absorption \emph{stamped}
by a clock transition, the stamp $e^{i\omega\hat\tau_k}$ and the energy
bookkeeping being one and the same operator.

\begin{theorem}[The square is conserved by time-stamped counting]
\label{thm:conserve}
Let $O$ be any of the nine observables of Eq.~\eqref{eq:square} [or of
Eq.~\eqref{eq:pcops}, or of Proposition~\ref{prop:digits}]. Then:
\begin{enumerate}
\item[(a)] \emph{(Joint measurability with the counts.)}
$[O,N_1]=[O,N_2]=0$; each context possesses a joint projective
measurement with $(N_1,N_2)$, and a nonselective measurement of any
context leaves the full counting distribution $p(n_1,n_2)$ of any state
unchanged.
\item[(b)] \emph{(Invariance under detection.)} $[O,V_k]=0$, and
moreover $[O,a_kT_k]=[O,a_k^\dagger T_k^\dagger]=0$ separately. Hence,
in the interaction picture with respect to $H_0$, the square observables
are constants of the detection dynamics and invariants of the counting
semigroup: for the master equation
$\dot\rho=\sum_k\gamma_k\bigl(L_k\rho L_k^\dagger
-\tfrac12\{L_k^\dagger L_k,\rho\}\bigr)$ one has
$\frac{d}{dt}\Tr(O\rho)=0$, and every individual jump preserves each
$O$: $L_kOL_k^\dagger=O\,L_kL_k^\dagger$. The witness correlators are
the same whether measured before, during, or after the accumulation of
the click record.
\item[(c)] \emph{(The stripped observables are not conserved.)} The
counts themselves obey $[N_k,V_k]\neq0$ (counting must proceed), and the
clock-only counterparts
$\zeta_k=(-1)^{\lfloor \HC{}_{,k}/\omega\rfloor}$ and
$S^{\mathrm{C}}_k=e^{i\pi \HC{}_{,k}/\omega}$ \emph{anticommute} with
$V_k$,
\[
\zeta_k V_k\zeta_k=-V_k,\qquad
S^{\mathrm{C}}_k V_k S^{\mathrm{C}\,\dagger}_k=-V_k ,
\]
so every detection event flips them: a square built on the clocks
alone, with the modes as spectators, is scrambled by the counting
process. Mode-only squares do not exist (Proposition~\ref{prop:nogo}).
\end{enumerate}
\end{theorem}

Part (c) answers the natural objection that two ideal clocks are two
canonical pairs outright, so PC could be run on the clocks and the light
ignored. It could---but that square neither involves the counts (its
outcomes are independent of the click record) nor survives them (its
observables anticommute with the detection coupling and are randomized
by every event). Energy conservation of time-stamped counting
superselects the \emph{joint} square: among the clock-plus-mode
structures, exactly the modular functions of the conserved station
energies $E_k$ are measurable alongside the ongoing counting, and these
are the ones in which the photon-number parities appear
[Eq.~\eqref{eq:factorize}]. The light is not a spectator; the witness
consumes the click record itself.

Two bookkeeping remarks. First, under free evolution the square
precesses rigidly,
$e^{iH_0t}X_ke^{-iH_0t}=\cos(\omega t)X_k-\sin(\omega t)Y_k$ with $Z_k$
invariant, so a PM square is mapped to a PM square and
Theorem~\ref{thm:square} holds at every instant; fixing measurement
phases means working in the interaction picture, as usual for Ramsey
interferometry. Second, conservation gives the $Z$ outcomes a memory
interpretation: $E_k$-conservation means the outcome $(-1)^{j_k+n_k}$ is
the same whenever it is read. Early in the counting the quanta sit in
the mode ($n_k$ large, clock bin $j_k$ still at its initial value);
late, they sit in the clock ($n_k$ small, $j_k$ advanced by the number
absorbed): the two readings disagree about where the quanta are, never
about the conserved parity. With the clock register prepared within a
single energy cell, $Z_k$ reads the parity of the initial photon number
at every stage of the counting---precisely the coherence information
that destructive counting without a clock discards---while the $X$- and
$Y$-type observables measure coherences between adjacent cells of the
conserved energy, i.e.\ the optical phase information written into the
clock by the stamps $e^{i\omega\hat\tau_k}$.

\paragraph{Operational protocol.} For concreteness, one run of the
experiment, with the context assigned to that run:
\begin{enumerate}
\item[(1)] Prepare the input light (e.g.\ coherent states); the clocks
need no preparation.
\item[(2)] Measure the three observables of the run's context at one
common time, in the interaction picture---before, during, or after the
counting record accumulates, the choice being immaterial by
Theorem~\ref{thm:conserve}: $Z$-type entries by jointly reading the
photon count $n_k$ and the clock bin $j_k$ and outputting
$(-1)^{j_k+n_k}$; $X$- and $Y$-type entries by Ramsey interferometry
between adjacent $E_k$-cells, conditioned on the count; the entries of
$R_3,C_3$ by the corresponding joint two-station eigenbasis measurement.
The counting marginals are undisturbed throughout.
\item[(3)] Multiply the three outcomes to obtain the context value;
estimate $\chi$ across runs. Quantum mechanics gives $\chi=6$
identically; any noncontextual assignment gives $\chi\le4$.
\end{enumerate}

\section{Coherent light}\label{sec:coherent}

Let the input be $\ket{\alpha_1}\otimes\ket{\alpha_2}$ with arbitrary
clock states $\rho_{\mathrm{C}}$. By Corollary~\ref{cor:twolayer} and
Theorem~\ref{thm:conserve}(a), in every one of the six contexts the
counting marginals are the same exact product of Poissonians,
$p(n_1,n_2)=e^{-|\alpha_1|^2-|\alpha_2|^2}
\frac{|\alpha_1|^{2n_1}}{n_1!}\frac{|\alpha_2|^{2n_2}}{n_2!}$, with
$Q_{\mathrm{Mandel}}=0$ and $g^{(2)}=1$: every counting-only functional
of the data is reproduced by a classical field. The same runs, read
jointly with the clock registers, return $\chi=6>4$. The deviation from
classicality announced in the title is carried by the very clicks that
look classical in isolation---concretely, by digit functions of (clock
bin $+$ photon count) and by cell coherences---and, by
Proposition~\ref{prop:nogo}, there is no way to carry it in the clicks
alone.

What is, and is not, certified deserves care; the following three
statements are logically independent and all three are true.
\begin{enumerate}
\item[(1)] Coherent states have a positive $P$ function; no
photodetection statistics of them, and more generally no
counting-compatible data of any state, witnesses contextuality
(Sec.~\ref{sec:nogo}).
\item[(2)] The clock-referenced observables violate noncontextuality
maximally and state-independently (Theorem~\ref{thm:square}). Being
state-independent, the violation is a property of the measured
observable algebra, not a nonclassicality certificate of the input state
\cite{BudroniRMP}---coherent light is the motivating input precisely
because statement (1) makes it the hardest case for counting, not
because the violation singles it out.
\item[(3)] Classical fields and classical simulations can reproduce
particular quantum correlation values when ``contexts'' are implemented
as classical measurements \cite{Frustaglia2016,Markiewicz2019}, and
contextuality experiments with intense or heralded light depend
delicately on how measurement events are defined
\cite{ZhangKim2019,Qu2021}. Our claims are intra-quantum: sharp,
repeatable projective contexts with compatibility relations fixed by the
operator algebra, the standard setting of state-independent
contextuality \cite{BudroniRMP}. The usual sharpness and compatibility
assumptions of that program are inherited, not resolved, here.
\end{enumerate}
Consistency with the equivalence between contextuality and Wigner
negativity for continuous-variable measurement models \cite{Booth2022}
is instructive: that equivalence concerns Gaussian-type measurements,
under which positive-Wigner states such as coherent states can never
reveal contextuality. Our contexts are far from Gaussian---spectral
projectors onto energy cells and conditioned clock translations---so the
negativity resides in the measurement operators, as it must for any
state-independent construction evaluated on Gaussian states
\cite{Asadian2015}.

\section{Physical clocks and finite registers}\label{sec:physclocks}

The ideal TQS has spectrum $\mathbb{R}$ and invites the usual objections
to unbounded-below Hamiltonians, addressed on their own terms in
Refs.~\cite{Stoica2024,Stoica2026,UnruhWald1989,HegerfeldtRuijsenaars1980}.
The construction, however, needs far less than the ideal clock.

\begin{proposition}[Cyclic registers are exact]\label{prop:cyclic}
Replace each clock by an $M$-level cyclic energy register with level
spacing $\delta=\omega/s$ ($s\in\mathbb{N}$ levels per photon quantum):
basis $\ket{m}$, $m\in\mathbb{Z}_M$, with $T=(\text{cyclic raise})^s$
and total energy index $e=m+sn \pmod M$. If $2s\mid M$, then all
statements of Lemma~\ref{lem:station}, Theorem~\ref{thm:square},
Proposition~\ref{prop:pcform}, Proposition~\ref{prop:digits} (for
$4s\mid M$), and Theorem~\ref{thm:conserve} hold \emph{exactly} on
$\mathbb{C}^M\otimes\mathcal{F}$, with $\hat\chi=6\,\id$ as an operator
identity. The minimal register is $M=2s$; for $s=1$, a single qubit per
detector whose flip operator is the time stamp.
\end{proposition}

The lattice relations are elementary (the key identity
$\lfloor(m+sn)/s\rfloor=\lfloor m/s\rfloor+n$ survives, and
$e^{i\pi e/s}$ is well defined on the cycle precisely when $2s\mid M$);
they are verified to machine precision in Appendix~\ref{app:numerics}.
Physically, an unbounded linear ladder (the idealization of a
Zeeman-type ladder, with integer spectrum unbounded in both directions)
realizes the same algebra with exact energy bookkeeping; the cyclic
register tracks energy modulo $M\delta$, which is immaterial for the
algebra and faithful whenever $M$ exceeds the relevant photon numbers.
The ideal TQS is the $M\to\infty$, $\delta\to0$ limit; no
unbounded-below spectrum is needed for any statement in this paper once
the cyclic register is adopted.

A hard-wall register (a truncated ladder of $K$ levels, the caricature
of a harmonic ancilla), by contrast, breaks the shift unitarity at its
edges and degrades the identity into an approximation controlled by the
register population near the walls. Table~\ref{tab:hardwall} shows
$\langle\hat\chi\rangle$ for two identical stations with coherent
inputs $|\alpha|^2=1.3$ and Gaussian register states of fixed energy
width $\sigma=3\delta$ centered mid-ladder: the noncontextual bound $4$
is cleared already at $K=8$, and $6-\langle\hat\chi\rangle$ falls
roughly like the Gaussian weight at the walls, below $10^{-13}$ by
$K=48$.

\begin{table}[t]
\centering
\begin{tabular}{c c c}
\hline\hline
$K$ (levels) & $\langle\hat\chi\rangle$ & $6-\langle\hat\chi\rangle$\\
\hline
 8 & 5.408983 & $5.9\times10^{-1}$\\
16 & 5.956500 & $4.4\times10^{-2}$\\
24 & 5.999365 & $6.4\times10^{-4}$\\
32 & 5.999998 & $1.6\times10^{-6}$\\
48 & 6.000000 & $4.6\times10^{-14}$\\
\hline\hline
\end{tabular}
\caption{Hard-wall (truncated-ladder) clock registers of $K$ levels,
$s=1$: $\langle\hat\chi\rangle$ for identical stations with coherent
inputs ($|\alpha|^2=1.3$, Fock cutoff $n_{\max}=12$) and Gaussian
register states of width $\sigma=3$ levels centered mid-ladder. The
noncontextual bound is $4$; the cyclic register of
Proposition~\ref{prop:cyclic} gives $6$ exactly at any size. Numerical
details in Appendix~\ref{app:numerics}.}
\label{tab:hardwall}
\end{table}

On the experimental side we restrict ourselves to the structural remarks
that the construction fixes. The coupling Eq.~\eqref{eq:Vk} is a
Jaynes--Cummings exchange with a qudit ladder (minimally a qubit) in
place of the atom, the configuration native to cavity and circuit QED;
$Z$-type contexts need number-resolved counting plus register readout in
the energy basis; $X$-type contexts need Ramsey interferometry on the
register conditioned on the count; and the joint contexts $R_3$, $C_3$
need two-register joint (Bell-type) measurements, which is where the
experimental difficulty concentrates, as in every PM implementation. A
quantitative feasibility study (losses, unsharp registers, finite
detection efficiency, POVM contexts) is beyond the scope of this paper
and flagged in Sec.~\ref{sec:discussion}.

\section{Discussion}\label{sec:discussion}

\subsection{Relation to prior work}

The square itself is Mermin's algebra \cite{Mermin1990}; its
continuous-variable dress is Plastino and Cabello's
\cite{PlastinoCabello2010}; the clock is Stoica's \cite{Stoica2024}. The
contribution of this paper is the combination and the theorems that make
it physics rather than relabeling: the abelian no-go for
counting-compatible observables (Proposition~\ref{prop:nogo}), which
says that no existing or future counting-only scheme can show
contextuality; the dilation lemma (Lemma~\ref{lem:weyl}), which locates
the missing conjugate in the station energy; and the conservation
theorem (Theorem~\ref{thm:conserve}), which says that the clock-dressed
square is exactly the contextual structure that time-stamped,
energy-conserving detection leaves intact---while clock-only squares are
scrambled by it and mode-only squares do not exist. To our knowledge the
pairing of a conjugate-time clock with a contextuality witness has not
appeared before: continuous-variable squares use quadrature or
displacement observables
\cite{PlastinoCabello2010,Asadian2015,LaversanneFinot2017}, pseudospin
constructions use number-shift operators \cite{Chen2002,McKeown2011},
and both classes fail counting compatibility, $[O,N_k]\neq0$. The
single-photon time--frequency pair of Fabre, Keller, and Milman
\cite{Fabre2022} is the closest physical neighbor---there, time and
frequency of one photon form a canonical pair on a restricted
subspace---whereas here the pair $(\hat\tau_k,E_k)$ lives on the full
Fock space of the mode, with the clock as an external, dynamical
reference. The quantum-clock literature
\cite{PageWootters1983,GLM2015,SmithAhmadi2019,HSL2021,Chataignier2024,Stoica2024}
supplies the kinematics we use but, so far as we could determine,
contains no contextuality statement; conversely, the contextuality
literature reviewed in Ref.~\cite{BudroniRMP} contains no clock.
(Leggett--Garg and ``temporal contextuality'' tests concern sequential
measurements in parameter time and are unrelated to a time
\emph{operator}.)

A classical local oscillator also converts counting into phase-sensitive
measurement (homodyning), but at the price of replacing the signal count
by quadrature data, $[x_\theta,N]\neq0$: the record that remains is not
the photon count. The clock is the fully quantum, energy-accounted
version of a phase reference that \emph{adds} conjugate information
while provably disturbing neither the counts
(Theorem~\ref{thm:conserve}a) nor being disturbed by them
(Theorem~\ref{thm:conserve}b).

\subsection{Assumptions and open problems}

The construction inherits the standard epistemology of state-independent
contextuality \cite{BudroniRMP}: contexts are sharp projective
measurements whose compatibility is fixed by the algebra, and the
ideal-clock version uses projective readout of continuous spectra, for
which the same idealizations as in Ref.~\cite{PlastinoCabello2010}
apply; the cyclic-register version of Proposition~\ref{prop:cyclic}
removes the continuum but keeps sharpness. Open problems we consider
worthwhile: (i) a POVM/unsharp-register analysis with noise thresholds
for the inequality $\chi\le4$; (ii) the joint contexts $R_3$, $C_3$
realized by local operations and shared entanglement, turning the square
into a two-station game; (iii) multi-qubit digit encodings of one
conserved energy (Proposition~\ref{prop:digits}) as a
counting-compatible register of many effective qubits; (iv) sequential
(Leggett--Garg-type) versions exploiting Theorem~\ref{thm:conserve}(b),
in which the same conserved square is measured repeatedly during the
counting process; and (v) the relativistic extension, in which the
station energies become generators of local time translations and the
clock becomes a quantum reference frame---the natural meeting point with
algebraic formulations of contextuality in field theory.

\subsection{Conclusion}

Photon counting of coherent light is classical twice over: its
statistics admit a classical field model, and---as we
proved---\emph{all} counting-compatible observables admit a
noncontextual model for every state. Attaching a Stoica clock to each
detector changes the second statement maximally: the conserved station
energy acquires a genuine conjugate time, the Plastino--Cabello square
transcribes onto the pairs $(\hat\tau_k,E_k)$, and nine observables that
commute with both photon counts---and with the detection dynamics
itself---violate noncontextuality at the algebraic maximum $\chi=6$ in
every state, coherent light and uninitialized clocks included, with the
counting record left exactly intact. The timed photon count deviates
from classicality where the untimed count provably cannot.
\appendix

\section{Proofs}\label{app:proofs}

\subsection{Proof of Lemma~\ref{lem:weyl}}
Work in the clock-energy representation,
$\mathcal{K}\cong L^2(\mathbb{R}_\varepsilon)$ with $\HC$ the
multiplication by $\varepsilon$ and $\hat\tau=i\,d/d\varepsilon$ (so
$[\hat\tau,\HC]=i$). Then
$\mathcal{H}_k\cong\bigoplus_{n\in\mathbb{N}}L^2(\mathbb{R}_\varepsilon)$
with $E=\varepsilon+\omega n$ on the $n$th summand. Define
$(W_0\psi)_n(e)=\psi_n(e-\omega n)$, a unitary of
$\bigoplus_nL^2(\mathbb{R})\cong L^2(\mathbb{R})\otimes\ell^2(\mathbb{N})$.
Since the shift is by a constant within each summand,
\[
W_0\,E\,W_0^\dagger = e\cdot\otimes\,\id,\qquad
W_0\,\hat\tau\,W_0^\dagger = i\,\tfrac{d}{de}\otimes\,\id,\qquad
W_0\,N\,W_0^\dagger = \id\otimes\mathrm{diag}(n).
\]
The pair $(i\,d/de,\;e\cdot)$ on $L^2(\mathbb{R})$ obeys
$[i\,d/de,\,e\cdot]=i$ and is carried to the Schr\"odinger pair
$(\hat q,\hat p)$ by a Fourier transform composed with the parity
operator; let $W$ be $W_0$ followed by this map (acting trivially on
$\ell^2(\mathbb{N})$). Equation~\eqref{eq:dilation} follows, and the
Weyl relations~\eqref{eq:weylrel} hold because they hold for
$(\hat q,\hat p)$ and are preserved under unitaries. Self-adjointness of
$E$ is immediate in the representation $e\cdot\otimes\id$. \qed

\subsection{Proof of Lemma~\ref{lem:station}}
From $[\hat\tau,\HC]=i$ and the centrality of the commutator,
$T\HC T^\dagger=\HC+[i\omega\hat\tau,\HC]=\HC-\omega$, while $T$
commutes with $N$; hence $TET^\dagger=E-\omega$ and, for any Borel $f$,
$Tf(E)T^\dagger=f(E-\omega)$. With $c(E-\omega)=c(E)-1$ this gives
$TP^{\pm}T^\dagger=P^{\mp}$, i.e.
\begin{equation}
TP^{\pm}=P^{\mp}T,\qquad T^{\dagger}P^{\pm}=P^{\mp}T^{\dagger}.
\label{eq:intertwine}
\end{equation}
Likewise $TST^\dagger=e^{i\pi(E-\omega)/\omega}=-S$, i.e.\ $ST=-TS$.
Using Eq.~\eqref{eq:intertwine} repeatedly:
$X^\dagger=P^{+}T^\dagger+P^{-}T=T^{\dagger}P^{-}+TP^{+}=X$;
\[
X^2=TP^{+}TP^{+}+TP^{+}T^{\dagger}P^{-}+T^{\dagger}P^{-}TP^{+}
+T^{\dagger}P^{-}T^{\dagger}P^{-}
=0+P^{-}+P^{+}+0=\id ;
\]
$ZX=(P^{+}-P^{-})X=T^{\dagger}P^{-}-TP^{+}$ and
$XZ=TP^{+}-T^{\dagger}P^{-}$, so $\{X,Z\}=0$. Then $Y=iXZ$ obeys
$Y^\dagger=-iZX=iXZ=Y$, $Y^2=-XZXZ=X^2Z^2=\id$,
$\{X,Y\}=i(X^2Z+XZX)=i(Z-Z)=0$, $\{Y,Z\}=i(XZ^2+ZXZ)=i(X-X)=0$, and
$XZY=i(XZ)^2=i\,(-X^2Z^2)=-i\,\id$, hence also $ZXY=+i\,\id$.
Counting compatibility: $P^{\pm},Z,S$ are functions of $E$ and
$[E,N]=0$; $T$ acts on the clock factor alone. The factorizations
\eqref{eq:factorize} follow from
$\lfloor(\varepsilon+\omega n)/\omega\rfloor
=\lfloor\varepsilon/\omega\rfloor+n$ for integer $n$. For the qubit
factorization, note $X$ maps $\mathrm{ran}\,P^{+}$ unitarily onto
$\mathrm{ran}\,P^{-}$; writing
$\mathcal{H}_k=\mathrm{ran}\,P^{+}\oplus\mathrm{ran}\,P^{-}$ and
conjugating by $\mathrm{diag}(\id,u^\dagger)$, where $u$ is that
unitary, gives $X\cong\sigma_x\otimes\id$, $Z\cong\sigma_z\otimes\id$ on
$\mathbb{C}^2\otimes\mathrm{ran}\,P^{+}$. \qed

\subsection{Proof of Theorem~\ref{thm:square}}
\emph{Context compatibility.} Entries of $R_1$, $R_2$, $C_1$, $C_2$
commute because operators of different stations commute and each entry
commutes with itself times a commuting factor. For $R_3$ and $C_3$,
use: if $O=O_1\otimes O_2$ and $O'=O'_1\otimes O'_2$ with
$O_kO'_k=\epsilon_k O'_kO_k$, $\epsilon_k=\pm1$, then
$OO'=\epsilon_1\epsilon_2\,O'O$. Every pair within
$R_3=\{X_1Z_2,\,Z_1X_2,\,Y_1Y_2\}$ and
$C_3=\{X_1X_2,\,Z_1Z_2,\,Y_1Y_2\}$ picks up
$\epsilon_1=\epsilon_2=-1$ from Lemma~\ref{lem:station}, hence
commutes.

\emph{Products.} With all cross-station factors commuting,
$R_1=X_1^2\otimes X_2^2=\id$, $R_2=Z_1^2\otimes Z_2^2=\id$,
$C_1=X_1^2\otimes Z_2^2=\id$, $C_2=Z_1^2\otimes X_2^2=\id$, while
$R_3=(X_1Z_1Y_1)\otimes(Z_2X_2Y_2)=(-i)(+i)\,\id=\id$ and
$C_3=(X_1Z_1Y_1)\otimes(X_2Z_2Y_2)=(-i)^2\,\id=-\id$, using
$XZY=-i\,\id$, $ZXY=+i\,\id$. Hence $\hat\chi=6\,\id$ and
$\langle\chi\rangle_\rho=6$ for every density operator on $\mathcal{H}$.

\emph{Noncontextual bound.} In a deterministic noncontextual model each
of the nine observables receives a value $v(O)\in\{\pm1\}$ independent
of context, and the value of each context product is the product of its
entries' values (the third entry of each context is, in quantum theory,
determined with certainty by the other two in every state, which is the
operational justification of the product rule for sharp measurements
\cite{Cabello2008,BudroniRMP}). The six context values $c_1,\dots,c_6$
then satisfy $\prod_ic_i=\prod_Ov(O)^2=+1$, whereas the quantum signs
$(+,+,+,+,+,-)$ have product $-1$; so at least one context value
differs from its quantum sign, giving
$\chi=c_1+\dots+c_5-c_6\le 6-2=4$. Convex mixtures of deterministic
assignments (stochastic noncontextual models) obey the same bound. \qed

\subsection{Proof of Proposition~\ref{prop:pcform}}
All station-1 operators commute with all station-2 operators, and
within one station the only nontrivial relation is $S_kT_k=-T_kS_k$
(Lemma~\ref{lem:station}). \emph{Compatibility}: by the two-sign rule
quoted in the proof of Theorem~\ref{thm:square}, each pair within each
row and column of Eq.~\eqref{eq:pcops} accumulates either zero or two
anticommutations; e.g.\ for $(\beta,\gamma)$: on station 1,
$S_1^\dagger$ vs $T_1S_1$ anticommute; on station 2, $S_2^\dagger$ vs
$T_2^\dagger S_2$ anticommute; product of signs $+1$. The remaining
pairs are checked identically. \emph{Products}: using
$T^\dagger S^\dagger TS=-\id$ and $TS^\dagger T^\dagger S=-\id$ per
station,
\[
ABC=T_1S_2T_1^\dagger S_2^\dagger=\id,\quad
abc=T_2^\dagger S_1T_2S_1^\dagger=\id,\quad
\alpha\beta\gamma=
\underbrace{(T_1^\dagger S_1^\dagger T_1S_1)}_{-\id}\otimes
\underbrace{(T_2S_2^\dagger T_2^\dagger S_2)}_{-\id}=\id,
\]
$Aa\alpha=T_1T_2^\dagger T_2T_1^\dagger=\id$,
$Bb\beta=S_2S_1S_1^\dagger S_2^\dagger=\id$, and
$Cc\gamma=(T_1^\dagger S_1^\dagger T_1S_1)\otimes
(S_2^\dagger T_2T_2^\dagger S_2)=-\id$.
\emph{Single-exponential form}: for central commutator,
$e^{A+B}=e^Ae^Be^{-[A,B]/2}$; with $A=i\omega\hat\tau$,
$B=i\pi E/\omega$ one has $[A,B]=-i\pi$, so
$e^{i(\omega\hat\tau+\pi E/\omega)}=iTS$ and
$e^{-i(\omega\hat\tau-\pi E/\omega)}=-iT^\dagger S$, whence
$\gamma=(iT_1S_1)(-iT_2^\dagger S_2)=T_1S_1T_2^\dagger S_2$. The
counting compatibility of all nine entries is inherited from
Lemma~\ref{lem:station}. \qed

\subsection{Proof of Theorem~\ref{thm:conserve}}
(a) Each context is a finite set of commuting self-adjoint operators,
each commuting with $N_1,N_2$; the joint spectral measure of the
enlarged commuting family $\{$context$\}\cup\{N_1,N_2\}$ is the required
joint measurement. If $\{\Pi_j\}$ is the PVM of a context,
$[\Pi_j,N_k]=0$ gives, for any bounded $f$,
$\Tr\bigl[f(N_1,N_2)\sum_j\Pi_j\rho\Pi_j\bigr]
=\Tr\bigl[\sum_j\Pi_jf(N_1,N_2)\Pi_j\rho\bigr]=\Tr[f(N_1,N_2)\rho]$.

(b) $[T_k,V_k]=0$ since $T_k$ commutes with $a_k$, $T_k$,
$T_k^\dagger$; $[f(E_k),V_k]=0$ since $[E_k,V_k]=0$; hence all of
$P_k^\pm,Z_k,S_k,X_k,Y_k$ commute with $V_k$, and cross-station
commutation is trivial, so $[O,V_k]=0$ for all nine entries of each
square. For the refined statement, split $V_k=v_-+v_+$ with
$v_-=\lambda_k a_kT_k$ (lowers $N_k$ by one) and $v_+=v_-^\dagger$
(raises it). Since $[O,N_k]=0$, the operator $[O,v_-]$ maps the $N_k=n$
eigenspace into the $N_k=n-1$ eigenspace, while $[O,v_+]$ maps it into
$N_k=n+1$; the two commutators live in different blocks of the
$N_k$-grading, so $[O,v_-]+[O,v_+]=0$ forces each to vanish. With
$L_k=\sqrt{\gamma_k}\,a_kT_k$ and $[O,L_k]=[O,L_k^\dagger]=0$:
$L_kOL_k^\dagger=OL_kL_k^\dagger$ (each jump preserves $O$), and in the
Lindblad equation
$\frac{d}{dt}\Tr(O\rho)
=\sum_k\gamma_k\Tr\bigl[(L_k^\dagger OL_k
-\tfrac12\{L_k^\dagger L_k,O\})\rho\bigr]=0$
because $L_k^\dagger OL_k=L_k^\dagger L_kO$ and
$[L_k^\dagger L_k,O]=0$.

(c) $[N_k,a_kT_k]=-a_kT_k\neq0$. For the clock-only parities: since
$\zeta_kT_k=-T_k\zeta_k$ (the shift by $\omega$ flips the clock-bin
parity) and $\zeta_k$ commutes with $a_k$,
$\zeta_kV_k\zeta_k=-V_k$; similarly
$S^{\mathrm{C}}_kT_kS^{\mathrm{C}\,\dagger}_k
=e^{i\pi(\HC{}_{,k}+\omega)/\omega}e^{-i\pi\HC{}_{,k}/\omega}\,T_k=-T_k$
gives $S^{\mathrm{C}}_kV_kS^{\mathrm{C}\,\dagger}_k=-V_k$. An
observable that anticommutes with the generator of a dynamics is not
conserved by it; in the jump unraveling, conjugation by a jump flips
its sign: $L_k\zeta_k=-\zeta_kL_k$. \qed

\subsection{Proof of Proposition~\ref{prop:digits}}
In the energy representation of Lemma~\ref{lem:weyl} write
$e=\omega(c+f)$ with $c=\lfloor e/\omega\rfloor\in\mathbb{Z}$ and
$f\in[0,1)$, so
$L^2(\mathbb{R}_e)\cong\ell^2(\mathbb{Z}_c)\otimes L^2([0,1)_f)$, and
decompose $c\mapsto(d_0,d_1,r)$ with $d_0=c\bmod2$,
$d_1=\lfloor c/2\rfloor\bmod2$, $r=\lfloor c/4\rfloor$, a bijection
$\mathbb{Z}\to\mathbb{Z}_2\times\mathbb{Z}_2\times\mathbb{Z}$ (the
floor convention handles negative $c$ uniformly). The shift $T$ sends
$c\mapsto c+1$ at fixed $(f,n)$. On $d_0=0$ ($c$ even) it sets $d_0=1$
with $\lfloor(c+1)/2\rfloor=\lfloor c/2\rfloor$, so $(d_1,r)$ fixed; on
$d_0=1$, $T^\dagger$ sends $c\mapsto c-1$ with
$\lfloor(c-1)/2\rfloor=\lfloor c/2\rfloor$, again $(d_1,r)$ fixed;
hence $X_A=TQ^{(0)}_0+T^\dagger Q^{(0)}_1=\sigma_x^{(d_0)}\otimes\id$
exactly. Likewise $T^{2}$ on $d_1=0$ sends $c\mapsto c+2$: $d_0$ is
untouched, $\lfloor c/2\rfloor\mapsto\lfloor c/2\rfloor+1$ flips $d_1$,
and with $\lfloor c/2\rfloor$ even,
$\lfloor(c+2)/4\rfloor=\lfloor c/4\rfloor$ fixes $r$; $T^{\dagger2}$ on
$d_1=1$ inverts it; so $X_B=\sigma_x^{(d_1)}\otimes\id$. By definition
$Z_A=\sigma_z^{(d_0)}$, $Z_B=\sigma_z^{(d_1)}$. Two commuting copies of
the Pauli algebra result, and the PM square over $(A,B)$ satisfies
$\hat\chi=6\,\id$ by the computation of Theorem~\ref{thm:square}. Every
operator used is a function of $E$ or a $T$-power times such a
function, so counting compatibility and conservation under $V$
(including each jump part separately, by the grading argument) follow
as in Theorem~\ref{thm:conserve}. \qed

\subsection{Proof sketch of Proposition~\ref{prop:cyclic}}
On $\mathbb{C}^M\otimes\mathcal{F}$ with basis $\ket{m}\otimes\ket{n}$
(register level $m$, photon number $n$), set $e=m+sn$ and define $T$ as
the $s$-fold cyclic raise on $m$, $S=\mathrm{diag}\,e^{i\pi e/s}$,
$c=\lfloor e/s\rfloor$, and $P^\pm,Z,X,Y$ as in Eq.~\eqref{eq:XYZ}.
Well-definedness on the cycle $m\sim m+M$ requires $e^{i\pi M/s}=1$ and
$(-1)^{\lfloor(e+M)/s\rfloor}=(-1)^{\lfloor e/s\rfloor}$, both
equivalent to $2s\mid M$ (the digit $d_1$ additionally requires
$4s\mid M$). The identity
$\lfloor(m+sn)/s\rfloor=\lfloor m/s\rfloor+n$ and the relations
$TP^\pm=P^\mp T$, $ST=-TS$,
$[E_{\mathrm{lattice}},V]=0$ (with
$V=a\otimes T+a^\dagger\otimes T^\dagger$, which shifts $e$ by $0$)
hold verbatim, and the proofs above go through unchanged. For $M=2s$
the register is two-dimensional. All identities are confirmed
numerically to machine precision (Appendix~\ref{app:numerics}). \qed

\section{Numerical validation}\label{app:numerics}

All operator identities claimed in the text were verified by exact
finite-dimensional computation on the cyclic-register models of
Proposition~\ref{prop:cyclic}, implemented in a short
\textsc{Python/NumPy} script
(\texttt{tqs\_counting\_square\_check.py}, included as an ancillary
file); 37 checks pass, none fail.

Stations $(M,s,n_{\max})=(8,2,3)$ and $(6,1,4)$ were used for the
two-station squares, and $(2,1,5)$ for the minimal qubit register; Fock
spaces are truncated at $n_{\max}$, which is exact for the tested
identities. Verified to at most $4\times10^{-15}$ in max-norm (most
identities exactly $0$): hermiticity, unitarity and involutivity of
$X,Y,Z$; the anticommutation relations of Lemma~\ref{lem:station} and
$TS+ST=0$; $[X,N]=[Z,N]=0$; pairwise commutation inside all six
contexts and the context products $(+1,+1,+1,+1,+1,-1)$ for the
dichotomic square [Eq.~\eqref{eq:square}], the PC-form square
[Eq.~\eqref{eq:pcops}], and the single-station digit square
(Proposition~\ref{prop:digits}, on $(M,s)=(8,1)$); the operator
identity $\hat\chi-6\,\id=0$ in all three cases; $[O,V]=0$ for all nine
observables of each square, together with
$\|[N_1,V]\|\approx1.73\neq0$ for the count and the exact
anticommutations $\{\zeta,V\}=\{S^{\mathrm{C}},V\}=0$ of the clock-only
parities [Theorem~\ref{thm:conserve}(c)]; and the invariance of the
photon-number distribution under a nonselective measurement of the
$Z_1Z_2$ context for a coherent $\otimes$ random-clock input. Direct
evaluation gives $\langle\hat\chi\rangle=6.000000000000$ for a random
pure state, for coherent inputs $\alpha_1=1.1$, $\alpha_2=0.7$ with
random clock states, and for the maximally mixed state.

For hard-wall registers (truncated ladders, Sec.~\ref{sec:physclocks}),
the six context-product operators factorize over the two identical
stations as $X^2\!\otimes\!X^2$, $Z^2\!\otimes\!Z^2$,
$(XZY)\!\otimes\!(ZXY)$, $X^2\!\otimes\!Z^2$, $Z^2\!\otimes\!X^2$,
$-(XZY)\!\otimes\!(XZY)$, and $\langle\hat\chi\rangle$ was evaluated on
$\bigl[\text{Gaussian register}(\sigma=3\delta)\otimes
\ket{\alpha}\bigr]^{\otimes2}$ with $|\alpha|^2=1.3$, $n_{\max}=12$,
giving Table~\ref{tab:hardwall}.

\end{document}